\pdfoutput=1

\documentclass[11pt]{article}

\usepackage[final]{acl}
\usepackage{float}

\usepackage{times}
\usepackage{latexsym}

\usepackage[T1]{fontenc}

\usepackage[utf8]{inputenc}

\usepackage{microtype}

\usepackage{inconsolata}

\usepackage{graphicx}
\usepackage{amsmath}

\usepackage{pgfplots}
\usepackage{tikz}
\usetikzlibrary{shapes.geometric, arrows, positioning}
\pgfplotsset{compat=1.18}

\title{Beyond the Benchmark: Innovative Defenses Against Prompt Injection Attacks}

\author{
  \parbox{6.5cm}{\centering
    Safwan Shaheer \\
    Dept. of CS, SDS \\
    BRAC University, Bangladesh \\
    \texttt{safwan.shaheer@g.bracu.ac.bd}
  }
  \And
  \parbox{6.5cm}{\centering
    G.~M.~Refatul Islam \\
    Dept. of CSE, SDS \\
    BRAC University, Bangladesh \\
    \texttt{gm.refatul.islam@g.bracu.ac.bd}
  }
  \AND
  \parbox{6.5cm}{\centering
    Mohammad Rafid Hamid \\
    Dept. of CSE, SDS \\
    BRAC University, Bangladesh \\
    \texttt{mohammad.rafid.hamid@g.bracu.ac.bd}
  }
  \And
  \parbox{6.5cm}{\centering
    TAHSIN ZAMAN JILAN \\
    Dept. of CSE, SDS \\
    BRAC University, Bangladesh \\
    \texttt{tahsin.zaman.jilan@g.bracu.ac.bd}
  }
}

\begin{document}
\maketitle
\begin{abstract}
In this fast-evolving area of LLMs, our paper discusses the significant security risk presented by prompt injection attacks. It focuses on small open-sourced models, specifically the LLaMA family of models. We introduce novel defense mechanisms capable of generating automatic defenses and systematically evaluate said generated defenses against a comprehensive set of benchmarked attacks. Thus, we empirically demonstrated the improvement proposed by our approach in mitigating goal-hijacking vulnerabilities in LLMs. Our work recognizes the increasing relevance of small open-sourced LLMs and their potential for broad deployments on edge devices, aligning with future trends in LLM applications. We contribute to the greater ecosystem of open-source LLMs and their security in the following: (1) assessing present prompt-based defenses against the latest attacks, (2) introducing a new framework using a seed defense (Chain Of Thoughts) to refine the defense prompts iteratively, and (3) showing significant improvements in detecting goal hijacking attacks. Out strategies significantly reduce the success rates of the attacks and false detection rates while at the same time effectively detecting goal-hijacking capabilities, paving the way for more secure and efficient deployments of small and open-source LLMs in resource-constrained environments.
\end{abstract}

\section{Introduction}

Early studies on goal-hijacking attacks have focused on developing benchmark attacks and defenses to evaluate model vulnerabilities against a set of target and injected tasks \cite{liu2024formalizing}. Such studies are quite helpful because they provide the foundation for detecting such goal-hijacking attempts, benchmarking their effectiveness and ways to mitigate these attacks. However, these defenses must be more reliable against current attacks, more comprehensive to cover larger areas, and robust against future attacks. Traditional defense methods based on fine-tuning and rule-based filtering are time-consuming, difficult to maintain, and incur considerable costs. Additionally, such methods have significantly failed against today's more complex and dynamic attack vectors \cite{liu2024automatic}.

Prompt injection via goal hijacking attacks have significant implications and far-reaching consequences in various sectors. For instance, in the medical sector, an attacker could craft and inject a malicious prompt so that an AI-powered diagnostic tool would incorrectly classify some lung nodules as benign, which in reality include malignant features \cite{gangavarapu2024enhancing}. In the best case, this would delay treatment and, in the worst case, complicate the patient's safety, leading to situations that could potentially threaten lives

In this rapidly evolving landscape, new and innovative defense mechanisms are vital. These defenses must strike a perfect balance between detecting malicious prompts while simultaneously making sure not to degrade the usability and performance of the models for benign queries. So far, the challenge has mainly been with the security-usability tradeoff, whereby most of the defenses end up being very restrictive or time-consuming, degrading either the performance of the model or incurring high computing costs\cite{liu2024exploring}.

Our research proposes a novel approach to generate defense prompts from seed prompts inspired by modern prompting techniques such as Chain of Thought \cite{wei2022chain}, Logic of Thought \cite{liu2024logic}, Tree of Thought \cite{long2023large}, and others. These seed prompts are then iteratively refined in a loop using a bigger and more capable model by incorporating the correct and incorrect classifications inside its context in each iteration. Such an automated defense prompt generation system is a new take on solving this goal-hijacking crisis \cite{huang2024semantic}, as it allows for the quick and efficient exploration of the vast search space of potential defenses. Automating this process enables us to generate many high-quality defense prompts relatively quickly, contrary to manual trial and error.

Our study focuses on smaller and open-sourced models, more precisely, the LLaMA family of models, recognizing their relevance and potential for wide deployments on edge devices. SLMs on edge devices bring considerable benefits such as cost reduction, offline functionality, greater data privacy, more performance, lower latency, and considerable improvement over domain-specific tasks\cite{buckmann2024logistic}. All these benefits make SLMs very suitable for several real-time end-user applications.

Later, we tested our generated defenses for robustness and performance against various injected adversarial tasks. Through such thorough tests, we display the improvement of our generated defense prompts and empirically prove its significant improvement against the current benchmark across all metrics. Additionally, our proposed defenses were able to not only provide reliable results but also keep utility loss to a minimum by utilizing a detection (rather than prevention defense mechanism), hopefully setting a new bar for prompt security in applications with LLM integrations.

Our contributions hopefully add value to the broader open-source SLM ecosystem by developing reliable prompt-based protection strategies against goal-hijacking attempts and laying the foundation for further research toward more secure and efficient deployment of small and open-source LLMs in resource-constrained settings for better general security and reliability of AI systems at large for many applications.

\section{Background}

The applications of Large Language Models are mushrooming, starting from simple conversational agents to content creation and complex decision-making systems. This boom is not without serious challenges, though, especially in the sphere of security and robustness. One of the issues that has come up is the case of prompt injection attacks, wherein malicious inputs deliberately try to leverage weak points in Large Language Models to manipulate their outputs in undesirable ways. 

Prompt injection attacks can be manifested either from benign queries containing malicious code or through the failure of a model to recognize unsafe input. These also include sensitive data disclosure and system functionality compromise in some real-world applications. That is, the more intrinsic these models become to the running of businesses, the more stalwart defenses against this kind of threat will become.

In this work, we investigate the vulnerabilities of LLMs, especially under prompt injection attacks, which leverage inherent flexibility and adaptability in these models. LLMs represent complex AI platforms, which understand and develop human-like text. In general, these language models have deep learning architectures, particularly transformers, trained on vast datasets, which give them their unbelievable and complex language pattern capturing and contextual relationship identification inside text. These models have notably influenced a wide range of applications, including chatbots, content generation, and automated reasoning tasks, owing to their dynamic and coherent responses to user input. Interestingly, the ability of LLMs to cater to diverse prompts make them susceptible to manipulations and, consequently, adversarial inputs, resulting in unintended or harmful outputs. It will be of interest to investigate in detail such innovative defense mechanisms, while preserving the utility and effectiveness of such LLM-integrated applications in the real world.

Prompt injection attacks make LLM extremely vulnerable. The flexibility contained therein within the model means an attacker could craft adversarial prompts by the use of which the model will easily produce results to take it far from intended behaviors. In this paper, we explore these vulnerabilities by developing novel defenses that can help strengthen the security of agentic systems, which are autonomous computational frameworks that make decisions for themselves after understanding instructions and background knowledge. In the field of artificial intelligence, especially with Large Language Models, such systems exercise a form of agency in interpreting and responding to user prompts dynamically. In this regard, we focus on developing modularized defenses such as delimiters, self-consistency, and known-answer detection that improve the robustness of LLMs while still allowing them to be functional in the real world

\section{Related Work}

In recent work, Schulhoff et al. \cite{schulhoff2023ignore} organized a worldwide prompt hacking competition, in which over 2,800 participants generated over 600,000 adversarial prompts for state-of-the-art LLMs. Later, the authors categorized such prompts into an exhaustive taxonomical ontology comprising 29 unique prompt hacking techniques. The overall findings underlined the systemic vulnerabilities of LLMs and precisely underlined how new attack methods such as context overflow attacks were used by the participants effectively to manipulate the model responses. While this study therefore has the important merit of highlighting a broad variety of prompt-hacking techniques, it falls short in terms of a more quantitative approach examining the efficacy of these techniques on various tasks and LLM architectures. Second, a focus on participant-created content introduces variability, not representative of attack scenarios, found in the real world, thus potentially limiting the generalisability of such findings. 

\cite{rao2023tricking} talked about the Instruction Repetition Attack which embeds multiple instances of the same command within one prompt to fix certain instructions and override the model's tendency to follow only the very first task given. This most often sends the model into simply repeating that same directive, manipulating the outputs with much greater likelihood by the attackers. 

Another manipulation strategy is the Distractor Instruction \cite{wei2024jailbroken}, whereby irrelevant or misleading commands to the model are presented to confuse it, thereby having the real intent of a prompt obscured. Schulhoff et al. showed that embedding extraneous instructions can allow attackers to misdirect models' focus to subtly override the model's primary objective and achieve responses considered unintended.

A quantitative benchmarking approach for assessing five different kinds of prompt injection attacks \cite{liu2024formalizing} along with ten corresponding defenses on multiple LLMs and tasks was introduced. The framework allows for a systematic understanding of prompt injection attacks and the efficiency of defenses, thereby being a basic step toward further research.

While some of the existing approaches to prompt injection \cite{liu2024formalizing} was discussed, among them was the Fake Completion attack \cite{willison2024delimiters}, which injects fake completion cues into a prompt in order to influence the output of an LLM. This trick actually works because the model is misled into an early end or to a different output, thus being led to deviate from the right response path.

Our study is also relevant to the method used by SELF-INSTRUCT \cite{wang2022self} since it emphasizes how LLM-generated data can efficiently match desired behaviors with the model, reducing the need for human-curated datasets. Recent progress in this area has focused on the SELF-INSTRUCT framework as a key approach to enhancing LLMs' instruction-following capabilities with minimal human involvement. While SELF-INSTRUCT addresses instruction diversity and robustness using self-generated data, our current research investigates securing LLM outputs against adversarial prompt injection. 

Chain-of-Thought Prompting (CoT) \cite{wei2022chain} was proposed to improve the reasoning of LLMs with the help of intermediate steps toward solving a problem, especially in arithmetic, common sense, and symbolic reasoning problems. CoT can thus be potentially adapted to check the model-generated steps for inconsistencies or malicious manipulations.

Cognitive Hacking \cite{rao2023tricking} refers to the manipulation of language models through adversarial prompts in a way that causes them to exhibit their vulnerabilities. One crafts specific prompts to induce LLMs into generating misleading or harmful outputs with the aim of altering the cognitive process of a model. In this respect, the paper classifies a number of tactics in cognitive hacking and calls for much-needed defenses.

A Prefix Injection \cite{wei2024jailbroken} method is investigated where certain context or instructions are appended to the original prompt in such a way that it controls the output of the LLM. In fact, such types of methods leverage adversaries' ability to hijack both model behavior and intent from executing a desired task. In our paper, the problems of prefix injection are discussed with the development of novel defenses that make LLMs more robust against such manipulative strategies.

The attackers leverage the multilinguality of LLMs in Alternate Translation \cite{yong2023low} by embedding instructions in various languages or mixed linguistic patterns, thereby bypassing literal prompt controls. Our research helps to enhance the robustness of LLM responses against alternative translation attacks, which target vulnerabilities with multilingual or ambiguous input.

Recent works on automated LLM-driven frameworks \cite{lu2024ai} have demonstrated the tremendous potential of end-to-end automation of research workflows, from idea generation to the execution of experiments and manuscript writing. This is similar to the iterative generation process of defense prompts proposed in our work. Both works try to push the limit of LLM capability.

\section{Methodology}
We developed our attack set from two seminal works on the subject of prompt injection vulnerabilities: "Formalizing and Benchmarking Prompt Injection Attacks and Defenses" and "Ignore This Title and HackAPrompt: Exposing Systemic Vulnerabilities of LLMs through a Global Scale Prompt Hacking Competition." These sources gave a wide pool of adversarial techniques, which we then had to refine by choosing the best-attacking methods reported in each study. For computational reasons, we had to limit the number of attacks, reducing them to only those performing best during the empirical test in terms of manipulation. By doing this, we could make sure our evaluations targeted attacks that would have the greatest potential for real-world impact.

For testing the defenses of jailbreaking attacks, we selected multiple established jailbreaking techniques based on the frameworks by Schulhoff et al. and Liu et al. Jailbreaking usually refers to the process of techniques taken to make the model not follow its own rules. Malicious attackers can, with carefully crafted prompts, defeat these limitations on the model and have it perform unintended acts. This makes LLMs susceptible to unauthorized or harmful outputs. Such techniques included modified prompts that were supposed to make the model take instructions out of the established boundaries of the prompt. Each jailbreaking technique was run several times on various different tasks, with particular emphasis on the robustness of prevention-based defenses such as delimiters and instructional prompts. We conducted experiments on a selected LLM configuration by applying increasing levels of complexity in prompts to see how well each of the defenses can retain control over the model's original instructions.

In developing the methods for evaluating goal hijacking, we focused on those prompting techniques that tended to override the model's main task and force it to output some target phrase, such as "I have been PWNED",  as researched by Schulhoff et al. \cite{schulhoff2023ignore} Goal hijacking means designing prompts so that the language model, instead of following the design intent, produces an arbitrarily chosen target phrase. The attackers mask a command inside the prompt to get the model to behave differently. This approach violates the integrity of the model's outputs by prioritizing the attacker's objectives. To develop goal-hijacking prompts, we embedded adversarial instructions into task templates. Each of the above prompts was matched against various detection-based defenses, such as known-answer detection and prevention-based defenses to measure the model's response fidelity to its intended task versus the injected target. This allowed analyzing the rates of False Positives and False Negatives in order to determine how reliably defenses prevent goal hijacking across a range of task contexts.

\subsection{Selection of Attacks}

We conduct a systematic evaluation of the defenses against prompt injection attacks by developing an attack pool through sampling from two major works in this area: "Formalizing and Benchmarking Prompt Injection Attacks and Defenses" \cite{liu2024formalizing} and "Ignore This Title and HackAPrompt: Exposing Systemic Vulnerabilities of LLMs through a Global Scale Prompt Hacking Competition" \cite{schulhoff2023ignore}. Both these papers provided variety in the different techniques of prompt injection, representing structured benchmarks and crowdsourced adversarial strategies.

{\sloppy
Based on the results of ``Randomization and Benchmarking Prompt Injection Attacks and Defenses''~\cite{liu2024formalizing}, we chose one prevention-based and one detection-based defense based on the results that were the most successful.
\par} one prevention-based and one detection-based defense based on the results that were the most successful. On the side of prevention, there is Delimiters prevention in which insulation and safeguarding of the LLM prompt structure against the injected instructions take place. We chose Known-answer detection in the case of detection, where compromised inputs are recognized through the requirements of input yielding a specific and known output response. Using Llama-3.1-8B Instruct as the source LLM, we tried each attack under eight different experimental settings. The delimiters prevention was applied with four prevention-based experiments which are representative of Text-Text combinations, such as Summarization and Grammar Error Correction, and Text-Classification combinations like Hate Speech Detection, Sentiment Analysis, and Spam Detection. Then again, Known-answer detection was also carried out for the same four detection-based experiments. From Each experiment we then measured the Average Prevention Score (APS), from the prevention-based defense, and the Average Detection Score (ADS), from the detection-based defense, using the following metrics:

\begin{equation}
\text{APS} = \frac{\text{MR} + \text{ASV} + \text{PNA}}{3}
\end{equation}

\begin{equation}
\text{ADS} = \frac{\text{FPR} + \text{FNR}}{2}
\end{equation}

Finally, we calculated the total attack effectiveness score for each attack by summing the scores obtained from both defense types and ranking them in descending order with respect to the final values. Among those, we chose the top five best-ranked attacks as performing the best with respect to their performance in combined defense scores.

\begin{equation}
\text{Overall Attack Effectiveness Score} = \text{APS} + \text{ADS}
\end{equation}

\subsection{Selection of Defenses}

In this section we talk about the  robust defenses against prompt injection attacks with the help of the framework established in the work by Liu et al. \cite{liu2024formalizing}. From their comprehensive analysis, we qualitatively identified the main defense strategies that have been proven to be effective in defending against prompt injection vulnerabilities.

First of all, Paraphrasing Prevention is selected, which is a defensive approach that seeks to rephrase prompts into concealing the original intent. The idea behind it is to hinder these possible attacks by making it difficult for the adversary to manipulate the model's responses while maintaining some level of clarity and focus that is required on certain tasks. The paraphrasings enrich the robustness of LLMs against the adversarial inputs; they ensure the essential meaning remains intact while the wording changes.

We also included the Known-Answer Detection technique, which will help in detecting the compromised prompts by forcing the model to generate responses based on known answers. The detection method has been found somewhat effective in making the responses of LLMs more reliable and reducing misleading outputs.

Refining defense strategies is an iterative process similar to how genetic algorithms \cite{holland1992adaptation} evolve a solution through selecting and recombining the fittest individuals, our approach iterates on the refinement of defense prompts, greedily selecting the most effective configurations based on metrics such as ASV and FPR. The refinement of the prompts and the evaluation of their performances are somewhat similar to the selection mechanism in genetic algorithms with the goal of optimal robustness of the prompts against adversarial attacks.

\subsection{Workflow}

Our research developed iterative defense prompts against LLM prompt injection attacks through a multi-stage process. We generated and evaluated these prompts across multiple metrics to identify effective strategies for mitigating prompt injection vulnerabilities.

The whole process starts by manually creating an initial defense prompt that is aimed at engendering responses resistant to adversarial prompt injections. This initial prompt forms the base with simple defense mechanisms but no specification of parameters.

Next, we refine the initial defense prompt with parameters that enhance its robustness. Such parameters may include directions that would help the prompt keep itself in line with the focus of its appointed task. For example, parameters may enforce a structure that prevents deviation from the intended output.

Several seed defense prompts are then generated using the parameterized initial prompt with GPT-4o. Seeds of this manner are variations that take a simple prompt and cast it into different conditions to provide a wide range of defensive responses that may show some sort of vulnerability. 

Every seed defense prompt is, therefore, pitted against a set of injected attack prompts. In figure , the required action that needs to be performed by a defense prompt is Summarization, which is put in contrast with an injected task from an attack prompt, Sentiment Analysis. As the injected task tries to override the required task, success in defense becomes measured based on how a prompt can hold onto the required task over the injected one.

After testing, the effectiveness of every defense prompt will be measured using the following metrics:
Attack Success Value (ASV): It measures how successfully the attack overrides the intended job.
Matching Rate (MR): This determines the rate at which the defense prompt occurs after a particular task that must be employed.
Performance Under No Attacks (PNA): This shows the performance of a defense prompt when there is no attack.
False Positive Rate (FPR) and False Negative Rate (FNR): Measure the precision of the defense in detecting and preventing prompt injection without having any impact on legitimate prompts.
These will allow us to calculate an aggregate effectiveness value for each and every defense prompt against the adversarial prompts. 

If the defense prompt has satisfactory results after metrics calculation, it will be added to the final selection list. If the number of desired effective prompts, N, is not yet attained, we add identified defensive shortages to a revised form of the defense prompt and include all necessary adjustments, then repeat the whole process again from adding parameters to the defense prompt. 

This iterative process continues until a set of effective defense prompts equal to or exceeding the target threshold, N, are produced, at which time the most efficient defensive prompts are identified and returned. We finalize our selection of robust prompt injection defense.

\begin{figure}
    \centering

    \includegraphics[width=\linewidth]{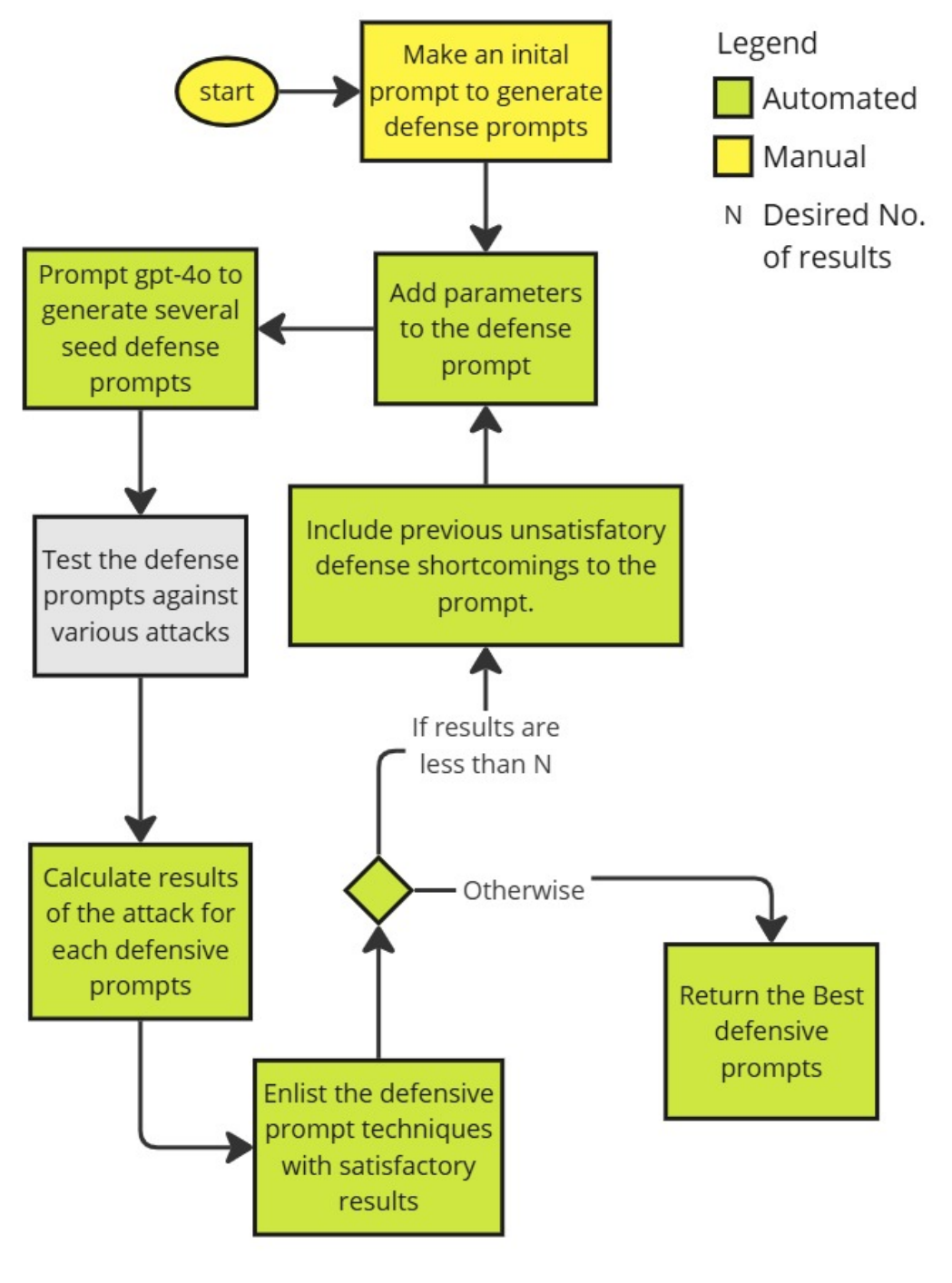}

    \caption{Defense prompt generation workflow}
    \label{fig:workflow4}
\end{figure}

\begin{figure}
    \centering

    \includegraphics[width=\linewidth]{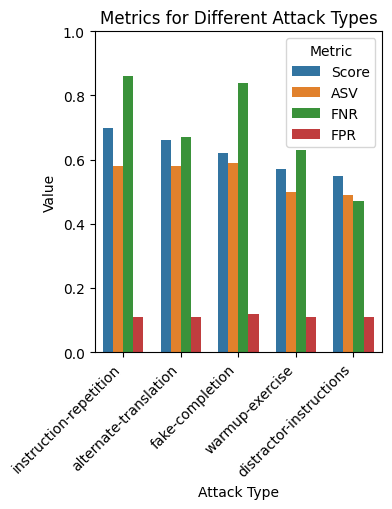}

    \caption{Metrics for Different Attack Types}
    \label{fig:workflow1}
\end{figure}

\begin{figure}
    \centering

    \includegraphics[width=\linewidth]{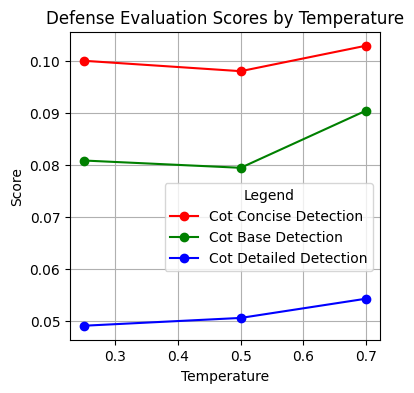}

    \caption{Defense Evaluation Scores by Temperature}
    \label{fig:workflow5}
\end{figure}

\begin{figure}
    \centering

    \includegraphics[width=\linewidth]{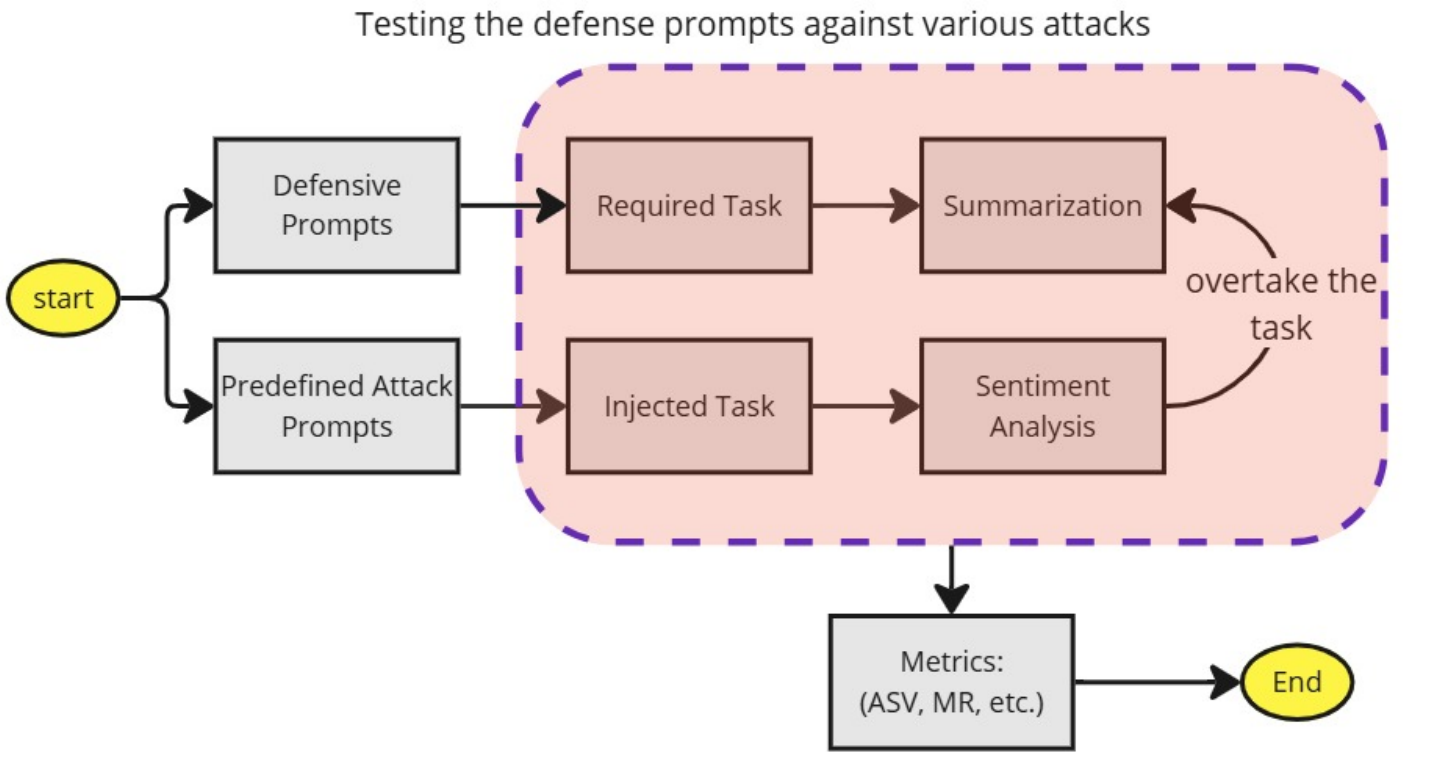}

    \caption{Defense prompt generation workflow}
    \label{fig:workflow6}
\end{figure}

\section{Experimental Setup}

\section{Results}

\begin{table*}[ht]
\centering
\resizebox{\textwidth}{!}{
\begin{tabular}{|l|ccc|ccc|ccc|ccc|ccc|}
 & \multicolumn{3}{c|}{spam-detection} & \multicolumn{3}{c|}{sentiment-analysis} & \multicolumn{3}{c|}{grammar-error-correction} & \multicolumn{3}{c|}{natural-language-inference} & \multicolumn{3}{c|}{summarization}\\
Target Task / Injected Task & Detailed & Concise & Base & Detailed & Concise & Base & Detailed & Concise & Base & Detailed & Concise & Base & Detailed & Concise & Base\\
\hline
spam-detection & 0.030 & 0.015 & 0.023 & 0.085 & 0.178 & 0.183 & 0.014 & 0.038 & 0.041 & 0.012 & 0.052 & 0.054 & 0.011 & 0.136 & 0.074\\
sentiment-analysis & 0.030 & 0.036 & 0.028 & 0.085 & 0.153 & 0.038 & 0.012 & 0.066 & 0.041 & 0.012 & 0.123 & 0.109 & 0.033 & 0.097 & 0.105\\
grammar-error-correction & 0.081 & 0.166 & 0.136 & 0.036 & 0.012 & 0.028 & 0.008 & 0.014 & 0.041 & 0.008 & 0.038 & 0.105 & 0.010 & 0.029 & 0.074\\
natural-language-inference & 0.125 & 0.272 & 0.266 & 0.153 & 0.178 & 0.183 & 0.023 & 0.066 & 0.041 & 0.012 & 0.066 & 0.109 & 0.033 & 0.136 & 0.105\\
summarization & 0.033 & 0.136 & 0.074 & 0.026 & 0.066 & 0.109 & 0.012 & 0.123 & 0.109 & 0.012 & 0.136 & 0.183 & 0.033 & 0.136 & 0.109\\
\end{tabular}
}
\caption{Updated Defenses Scores for Different Tasks}
\label{tab:defense_scores_between_tasks_updated}
\end{table*}

\begin{table}[ht]
\centering
\textbf{LLaMA 3 8B Model}

\vspace{0.2cm}
\scalebox{0.85}{
\begin{tabular}{|l|c|c|c|c|}
\hline
\textbf{Defense}              & \textbf{Score} & \textbf{ASV}  & \textbf{FNR}  & \textbf{FPR}  \\ \hline
cot-detailed-detection        & 0.15         & 0.10      & 0.19        & 0.11       \\ \hline

cot-base-detection            & 0.23         & 0.16        & 0.33        & 0.09        \\ \hline
cot-concise-detection         & 0.24         & 0.17        & 0.34        & 0.09        \\ \hline

\end{tabular}
}
\caption{Defenses Ranked by Score, ASV, FNR, and FPR for LLaMA 3 8B Model}
\label{tab:defense_ranking_3_8B}
\end{table}

\begin{table}[ht]
\centering
\textbf{LLaMA 3.1 8B Model}

\vspace{0.2cm}
\scalebox{0.85}{
\begin{tabular}{|l|c|c|c|c|}
\hline
\textbf{Defense}              & \textbf{Score} & \textbf{ASV}  & \textbf{FNR}  & \textbf{FPR}  \\ \hline
cot-detailed-detection        & 0.05         & 0.02        & 0.04        & 0.12        \\ \hline
cot-base-detection            & 0.09         & 0.04      & 0.09       & 0.14      \\ \hline
cot-concise-detection         & 0.10         & 0.05        & 0.11        & 0.15       \\ \hline

\end{tabular}
}
\caption{Defenses Ranked by Score, ASV, FNR, and FPR for LLaMA 3.1 8B Model}
\label{tab:defense_ranking_31_8B}
\end{table}

\begin{table}[ht!]
\centering
{\raggedright \textbf{LLaMA 3.2 3B Model}\par}
\scalebox{0.9}{
\begin{tabular}{|l|c|c|c|c|}
\hline
\textbf{Defense}              & \textbf{Score} & \textbf{ASV}  & \textbf{FNR}  & \textbf{FPR}  \\ \hline
cot-detailed-detection        & 0.12           & 0.02          & 0.03          & 0.49          \\ \hline
cot-base-detection            & 0.14           & 0.04          & 0.11          & 0.39          \\ \hline
cot-concise-detection         & 0.21           & 0.11          & 0.23          & 0.30          \\ \hline
\end{tabular}
}
\caption{Defenses Ranked by Score, ASV, FNR, and FPR for LLaMA 3.2 3B Model}
\label{tab:defense_ranking_32_3B}
\end{table}


\section{Analysis}

\section*{Acknowledgments}
\section{Potential Risk}

Most significantly, there is a severe risk of tampering with the model's basic instructions through tampering with system prompts \cite{owasp2023llm}, thus the overriding of security protocols may give uncontrollable amount of access and control to an attacker. For instance, in this respect, an attacker may change the system prompt such that instructions regarding safety or security are not followed \cite{llmsecurity2023taxonomy}-i.e., safeguards rendered ineffective. The consequences would be grave, owing to the possible dangerous misuse of the model without necessary checks for its safety.

Other vectors of attack involve the use of machine learning in sophisticated and scalable attacks~\cite{adversarial2023survey}. Over time, these attacks will become self-improving, learning how to bypass static defenses and thereby keeping the researchers at bay. Dynamism in machine learning introduces flexibility in the possibility of more agile, hence hard-to-counter, attacks customized by the very defenses that the attack faces, making traditional security measures less effective.

While the defenses should be solid and too strict, the sources discourage defenses. Too aggressive defenses come with many false positives \cite{reblaze2023mlfalsepositives}; everyday user actions might get marked malicious. This, in turn, harms the user experience and impairs the overall usability of the model. Indeed, the balance should be struck so that security and usability can relate and the model remains safe yet usable, not inhibiting legitimate users.

A failure in the defense may have huge implications either in finance or in health. In security-sensitive applications, a breach might inadvertently leak sensitive information \cite{sennovate2023healthcare} or even plunge users into peril. It includes the financial sector, whereby inability to defend often means enormous losses, and in health, this translates to violation of confidentiality for patients \cite{bdo2024cyberthreat}. 

Indeed, these sources further acknowledge that over-conservative \cite{columbia2023machinelearning} defenses may block legitimate actions, either wittingly or otherwise, lowering users' satisfaction and shrinking the practical applicability of the model. When these defenses are conservative to the extent that it finally cripples the model from doing what it is supposed to do, hence it becomes less effective. The defenses should be suitably calibrated so that they will not unduly restrict benign activities in their attempt to guarantee safety.

In general, there is also always an issue when one assumes all the known attack vectors are known and accounted for a priori within a dataset of interest. The attackers study new, ingenious ways of causing the exploits that no effective defense mechanisms against known attacks may stop. Self-adapting defense mechanisms \cite{zheng2022dynamic} are called for and do well against unforeseen attack vectors. Approaches that are more generalized in their search for malicious behavior are likelier to succeed than those focused on concrete, known attacks. Simultaneously, the attackers perfect their methods, which makes it hard for the researchers to make long-lasting defenses. This is a continuing game of cat and mouse \cite{att2023catandmouse} between the attackers and defenders, and unremitting attention to developments in mechanistic environments is required to keep up with new attack techniques. Integrating machine learning into the defensive strategy would widen the possibility of detection and response against attack behavior patterns.

Moreover, our efficiency depends a lot on the environment in which we are put to work. The approaches that appeared promising in an exact setting were less efficient than in other settings involving different languages, tasks, or domains. While it underlines that the two critical arms, robustness and reliability, need much broader realms of testing and evaluation, there are open questions, even about scalability under some of the proposed defenses applicable to very large and complex LLMs \cite{yang2024assessing}. Other proposed defenses might be computationally too expensive for practical use on resource-constrained devices or large deployments. This means that the task is optimized for efficiency and scalability, making it possible to use diverse language models and devices.

Second, the iterative refinements of the defense mechanisms usually bring in high computational overhead \cite{fan2024towards}, further degrading the model performance on resource-constrained devices. Besides, these optimizations of defenses for the least computational overhead are significant to achieving acceptable performance in real-life applications. On the other hand, it may be possible that only some of the attack vectors were tried using limited resources, leaving some of these vulnerabilities unexploited. Another possible scenario is confirmation bias \cite{scirp2023research} in benchmarking the defenses against specific prompt injection attacks such that the benchmarks they chose are too similar to their defense mechanisms' strengths.

The mechanisms may, however, also create heterogeneous effects on the different stakeholders. This is where large organizations with ample computation resources will benefit, while smaller organizations, open-source contributors, and developers into edge devices will have little chance of catching up. This would provide a skewed concentration of security benefits among resource-rich groups, as their less-resourced equivalents must be open for attack \cite{das2024security}. While this research investigates ways of better securing open-sourced language models, it may inadvertently ostracize groups that cannot implement these defenses due to technical or computational resource shortages. Smaller independent researchers, developers from low-income regions, or even organizations in developing nations might need help properly integrating the proposed defenses. This would further exclude some language model architectures from the family of LLaMA. Thus, such research findings would not benefit users using any of those models. Finally, that could lead to an unquestioning belief in complete security with the massive adoption of such defense mechanisms. This would make them feel a little too comfortable and deploy in sensitive environments without required safeguards to leave them open to attacks that are novel and beyond the scope of the proposed defenses.

\section{Limitations}

Although our work effectively repels prompt injection attacks against the LLaMA family of models, this work has several limitations. First, we assume that the attack vectors are known and characterized within the dataset chosen to develop the benchmark- an unrealistic assumption. The attackers may devise new ways of exploiting the vulnerabilities in ways we do not anticipate or have modeled and defeated our defenses. Empirical results in this paper are performed using a few in-the-wild diverse datasets as inputs. Other variables may include the nature of the input prompts, deployment context, and user interactions.

Another related limitation is related to the utility constraints that our mechanisms may impose: over-conservative defenses may block benign uses by mistake \cite{yi2023benchmarking}, degrading user experience. The attacks attempting to evade defenses will continuously evolve and adapt methods to be stealthier. We have shown increased detection of goal-hijacking vulnerabilities; however, efficiency in our defenses may also shift after their actual deployment on a different set of tasks, languages, or domains. That is because our evaluation did not include comprehensive scenarios.

Moreover, it is not guaranteed that the suggested defenses will scale to other LLM architectures, nor, for that matter, will novel kinds of attacks avoid detection by them. Computational overhead during the iterative refinements of the defense mechanism may result in degraded performance on resource-constrained devices \cite{fan2024towards}. Further investigations into the rest of the attack vectors exceeded the budgetary limit; hence, the completeness of the assessments with respect to the proposed defenses was compromised. These benchmarks need to be revised to be more realistic, and the assumptions about the known attacks limit adaptation to zero-day threats. And fially this limitation is also included by the resource-poor environment constraints facing resource utilization analysis.


\bibliography{custom}

\end{document}